\documentclass[12pt,preprint]{aastex}

\received{}
\accepted{}
\journalid{}{}
\articleid{}{}
\paperid{}
\cpright{AAS}{2004}
\ccc{}

\slugcomment{Resubmitted to the {\it Astrophysical Journal}}

\shorttitle{Molecular line opacity of LiCl in the mid-IR spectra of BDs}
\shortauthors{Weck {\em et al.}}

\begin{document}

\title{Molecular line opacity of LiCl in the mid-infrared spectra of brown dwarfs }

\author{P. F. Weck}
\affil{Department of Chemistry, University of Nevada Las Vegas, \\
 4505 Maryland Parkway, Las Vegas, NV 89154-4003}
\email{weckp@unlv.nevada.edu}

\author{A. Schweitzer}
\affil{Hamburger Sternwarte, Universitaet Hamburg, \\
       Gojenbergsweg, D-21029 Hamburg, Germany}
\email{Andreas.Schweitzer@hs.uni-hamburg.de}

\author{K. Kirby}
\affil{Institute for Theoretical Atomic, Molecular \& Optical Physics,\\ 
 Harvard-Smithsonian Center for Astrophysics,
       60 Garden Street, Cambridge, MA 02138}
\email{kkirby@cfa.harvard.edu}

\author{P. H. Hauschildt}
\affil{Hamburger Sternwarte, Universitaet Hamburg, \\
       Gojenbergsweg, D-21029 Hamburg, Germany}
\email{yeti@hs.uni-hamburg.de}

\author{P. C. Stancil}
\affil{Department of Physics and Astronomy and Center for Simulational Physics, \\
The University of Georgia, Athens, GA 30602-2451}
\email{stancil@physast.uga.edu}


\begin{abstract}

We present a complete line list for the $X~^1\Sigma^+$ electronic 
ground state of $^7$Li$^{35}$Cl computed using fully quantum-mechanical techniques. 
This list includes transition energies and oscillator strengths in the spectral 
region $0.3-39,640.7~\mbox{cm}^{-1}$ for all allowed rovibrational transitions 
in absorption within the electronic ground state. 
The calculations were performed using an accurate hybrid potential 
constructed from a spectral inversion fit of experimental data and from recent multi-reference single- 
and double-excitation configuration interaction calculations. The line list was 
incorporated into the stellar atmosphere code {\tt PHOENIX} to compute spectra 
for a range of young to old T~dwarf models. 
The possibility of observing a signature of LiCl in absorption near 15.8 $\mu$m 
is addressed and the proposal to use this feature to estimate the total lithium 
elemental abundance 
for these cool objects is discussed.

\end{abstract}


\section{Introduction}

In the study of low-mass stellar objects, the presence or absence of the Li I 6708~\AA~
resonance line has played an important role in ascertaining whether the object is a 
brown dwarf. However, the use of this so-called {\it lithium test} \citep{reb92} to 
determine substellarity has some drawbacks. L dwarfs which lie just below the
bottom or at the edge of the hydrogen-burning main-sequence may have some period in 
their early evolution of lithium burning, depleting their lithium abundance, 
decreasing the strength of the Li I resonance line, and thereby suggesting they 
are main-sequence stars. 
Furthermore, the depletion of lithium is age dependent, which in turn can be used 
as a clock under the correct conditions \citep[see e.g.][and 
references therein]{bas96,cha97,bar99}.
The reduction of the strength of the Li I resonance line can also occur in lower 
temperature objects ($T_{\rm eff} < 1500$~K), near the L/T dwarf interface,
due to the sequestering of lithium into 
molecular species such as LiCl, LiH, and LiOH \citep{lod99}. 
In either case, conclusions drawn from the lithium test alone (like age determination, or 
substellarity in the case of L~dwarfs) may be inaccurate. 
   
Thermochemical equilibrium calculations of cool dwarf 
atmospheres \citep{lod99} suggest that LiCl is the dominant Li-bearing gas 
over an extended domain of the temperature-pressure diagram. LiCl has a large dipole 
moment in its ground electronic state
which may give rise to an intense rovibrational line spectrum in the 
mid-infrared near 15.8 $\mu$m. As such, LiCl may give a significant absorption 
feature in L and T dwarf spectra as suggested by \citet{lod99} and \citet{bur00}. If the feature 
is observable, it could be used to estimate the total lithium elemental abundance in conjunction 
with optical Li I observations, to confirm the equilibrium lithium chemistry models, and to provide 
a better test of substellarity for cool objects. 

In this work, we continue our long term project to update and complete molecular 
opacity data \citep*{wec03a,wec03b,wec03c,wec04}.
Here we present a complete line list 
(transition energies and oscillator strengths) of all allowed rovibrational 
transitions in the electronic ground state of $^7$Li$^{35}$Cl. The calculations
were performed using an accurate hybrid potential and the dipole moment function
of \citet{wec04}. The line list was incorporated into the stellar atmosphere code 
{\tt PHOENIX} \citep{hau99} to compute spectra for a range of T~dwarf 
models to explore the possibility of observing LiCl. 

\section{Molecular calculations}

For the present calculations, an accurate hybrid 
potential was constructed for the $X~^1\Sigma^+$ electronic 
state from the spectral inversion fit of \citet{ogi92} and from the 
multi-reference single- and double-excitation configuration interaction 
(MRSDCI) calculations of \citet{wec04}. 
The fit to the effective potential energy proposed by \citet{ogi92} 
consisted of a sum of five radial functions accounting empirically 
for vibrational adiabatic and nonadiabatic effects.
The coefficients of this expansion were determined by direct spectral 
inversion from the frequencies of 2577 known transitions in the 
infrared and microwave spectral regions for the isotopic variants 
$^6$Li$^{35}$Cl, $^6$Li$^{37}$Cl, $^7$Li$^{35}$Cl and $^7$Li$^{37}$Cl. 
The normalized standard deviation of the fit was 0.993 over the 
complete domain of definition of the radial functions, i.e. for internuclear 
distances from $R=3.25$ to $4.80~a_0$. 
A shift in energy of $-467.209627~\mbox{a.u.}$ was applied to the 
Ogilvie fit, in addition to a shift of $+0.0138~a_0$ from its original 
equilibrium geometry, to obtain coincidence with the {\em ab initio} energy 
minimum at $R_e=3.8185~a_0$ determined by cubic spline interpolation from the 
MRSDCI data of \citet{wec04}. Beyond the range $3.25 \leqslant R \leqslant 4.80~a_0$, a 
spline fit to the {\em ab initio} data was used, connecting smoothly 
with the shifted Ogilvie fit.
For internuclear distances 
$R> 50.0~\mbox{a}_0$, a fit to the multi-reference potential has been 
performed using the usual van der Waals dispersion expansion to account 
for the long-range interaction.
To our knowledge, no data have been reported for the van der Waals 
coefficients of the $X~^1\Sigma^+$ state of LiCl, thus theoretical 
estimates were obtained using average values from several techniques 
in a similar way as in \citet{wec03c}.

In order to determine the spectroscopic constants of the 
$X~^1\Sigma^+$ potential, the vibrational wave 
functions, $\chi_{v}(R)$, and energy eigenvalues, $G(v)$, have been  
calculated by solving with Numerov techniques \citep{coo61} 
the radial nuclear Schr\"{o}dinger equation,
\begin{equation}\label{re}
\left[ -\frac{1}{2\mu}\frac{d^2}{dR^2} 
+ E_{el}(R)+\frac{J(J+1)}{2\mu R^2}-G(v)\right] \chi_{v}(R)=0,
\end{equation} 
where $\mu$ is the reduced mass of the system, $J$ is the rotational quantum 
number corresponding to the angular momentum of nuclear rotation, 
and $E_{el}(R)$ is the electronic potential energy. 
The reduced mass adopted for $^{7}$Li$^{35}$Cl was $5.8435744~\mbox{u}$\footnote{In 
atomic mass units, Aston's scale} $=10,651.3431~\mbox{a.u.}$ \citep{hub79}. 
Calculations were performed on a grid with stepsize $1\times10^{-3}~a_0$ 
for the integration, over a range of internuclear distances from $R=2.5~a_0$ 
to $100.0~a_0$.

The calculations yielded for this hybrid potential an energy difference 
$D_e=G(v_{max})=G(141)=4.962~\mbox{eV}=0.182~\mbox{a.u.}$ 
and a dissociation energy  
$D_0=4.922~\mbox{eV}=0.181~\mbox{a.u.}$, slightly larger than 
the thermochemical value, $D_0=4.85~\mbox{eV}$, of \citet{bre61} 
or the flame photometry measurement, $D_0=4.79~\mbox{eV}$, 
of \citet{bul61}. The introduction of the spin-orbit 
interaction in our calculations would lower the 
dissociation asymptote of the $X~^1\Sigma^+$ state, thereby 
bringing the theoretical dissociation energy into closer 
agreement with the experimental estimates.
Our theoretical vibrational constants $\omega_e=642.97~\mbox{cm}^{-1}$, 
$\omega_ex_e=4.49~\mbox{cm}^{-1}$ and $\omega_ey_e=0.02~\mbox{cm}^{-1}$ 
are in excellent agreement with the accurate 
Dunham constants $Y_{10}=642.95813~\mbox{cm}^{-1}$, 
$-Y_{20}=4.475085~\mbox{cm}^{-1}$ and $Y_{30}=0.0208072~\mbox{cm}^{-1}$, 
respectively, derived from experiment by \citet{tho87} 
for $^{7}$Li$^{35}$Cl. The frequency of the first band, 
$\nu(1-0)=\Delta G_{1/2}=634.06~\mbox{cm}^{-1}$, essentially 
reproduces the value of $634.076~\mbox{cm}^{-1}$, obtained 
using the Dunham terms given above.  

The line oscillator strengths, $f^{ab}_{v'J',v''J''}$, from rovibrational 
states $v''J''$ to final state $v'J'$ were computed using the $X~^1\Sigma^+$ 
dipole moment function of \citet{wec04} for all allowed absorption transitions 
between the 29,370 rovibrational levels solutions of 
Eq. (\ref{re}), thus giving a total of 3,357,811 lines\footnote{The complete 
list of $^7$Li$^{35}$Cl oscillator strength data
is available online at the UGA Molecular Opacity Project database website
\url{http://www.physast.uga.edu/ugamop/}}. A standard 
expression for the line oscillator strength can be found, for example, in 
\citet{sko03}. Computed line oscillator strengths and transition 
energies are reported in Table \ref{tbl1}, along with the high resolution 
measurements of \citet{jon87}, for the $R$-branch $(\Delta J=J'-J''=+1)$ of 
the fundamental vibrational band $(\Delta v=v'-v''=+1)$, for $v''=0$ to 4. 
The agreement is excellent, with a maximum transition energy discrepancy 
of $\sim 0.32~\mbox{cm}^{-1}$ for the $R(58)$ line of the 
$1-0$ band.

\begin{figure}
\plotone{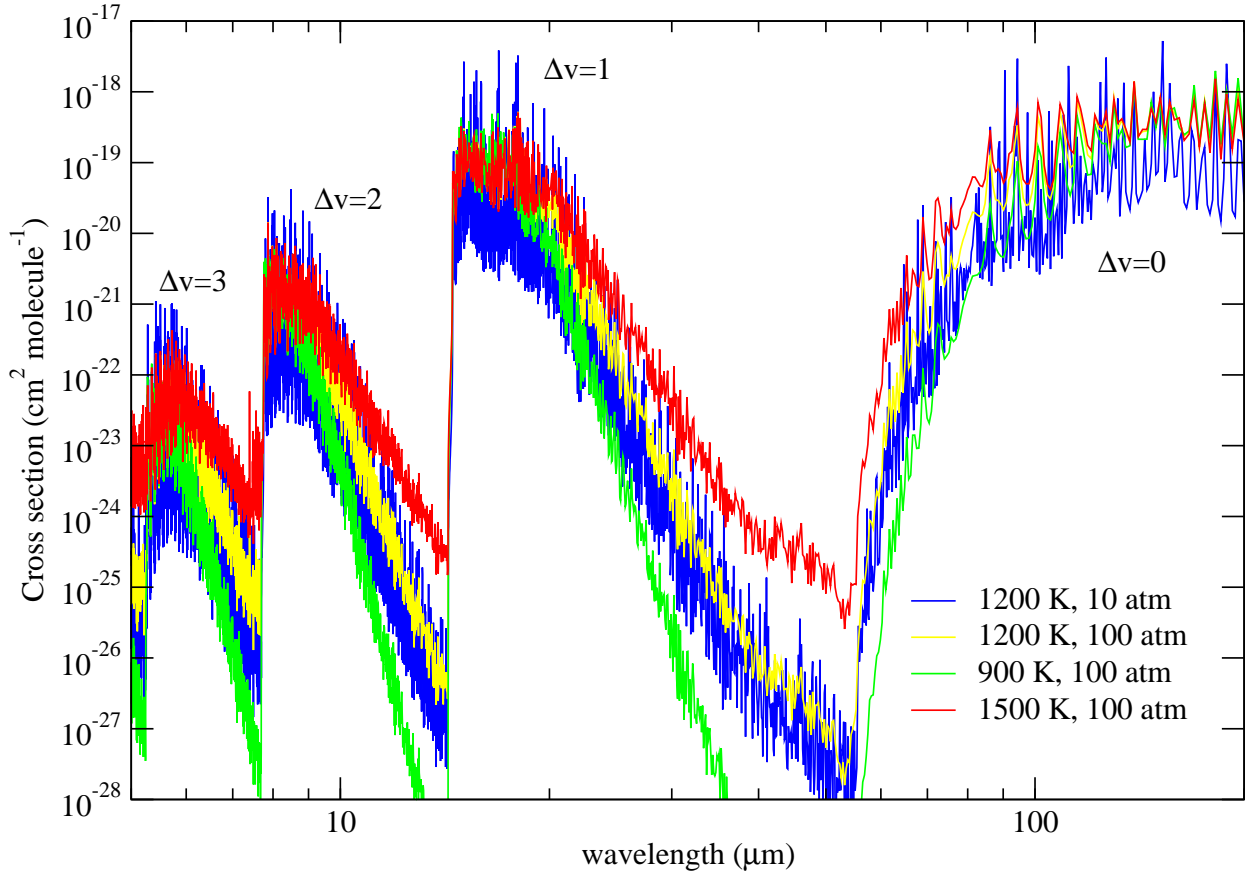}
\caption{Representative absorption cross sections for the $X~^1\Sigma^+$ state
of LiCl. 1200~K and 10 atm (blue). 900 (green), 1200 (yellow), and 
1500~K (red), all at 100 atm. The opacity for the fundamental and
first two overtone vibrational bands as well as a portion of the pure 
rotational band are shown. \label{fig1}}
\end{figure}

In Figure \ref{fig1}, representative LiCl opacities, absorption cross section 
per molecule, are presented for pressures and temperature appropriate to
T dwarfs. The opacities are computed using Eq.~(6) of \citet{dul03},
using Einstein A-coefficients from the above line lists,  
and multiplying by a Lorentzian line profile. The full width half
maximum line width is estimated by only considering collisional
broadening and is typically $\sim$0.1 cm$^{-1}$ at 100 atm. The
rovibrational levels of LiCl are assumed to be in equilibrium and
a correction for stimulated emission is included.  The fundamental
and first two vibrational overtone bands, as well as a portion of the
pure rotational band, are depicted. The fundamental band, with a
band origin at 15.8~$\mu$m, is the dominant LiCl opacity source in
the mid-infrared. We note, however, that PHOENIX uses molecular line
lists instead of pre-computed opacity tables as described below.


\section{{\tt PHOENIX} synthetic spectra}

The atmosphere models used for this work were calculated as described in
\cite{LimDust}. These models and their comparisons to earlier versions were
the subject of a previous publication \cite[]{LimDust} and we thus do not
repeat the detailed description of the models here. However, we will briefly
summarize the major physical properties. The models are based on the Ames
H$_2$O and TiO line lists by \cite{ames-water-new} and \cite{ames-tio} and
also include the line lists for FeH by \cite{FeHberk2} and for
VO and CrH by R. Freedman (NASA-Ames, private communication). 
We try as much as possible to constantly add new opacities as they become 
available \citep[see for example,][]{wec03a,wec03b} and the new FeH and CrH 
opacities recently calculated in \cite{bur02} and \cite{dul03} will soon 
be added to our database. However, as can be seen from these references, 
the new line lists calculated for FeH and CrH have no features (for 
vibrational transitions) in the mid-IR region where the LiCl feature is 
located. Although the global opacity is expected to be changed overall 
using these new line lists, for the purpose and wavelength window of 
this paper the use of the line lists of \cite{FeHberk2} and R. Freedman  
is appropriate.
The models
account for equilibrium formation of dust and condensates and include grain
opacities for 40 species. 
In this paper we only consider the so-called ``AMES-cond'' models in which the 
dust particles have sunk below the atmosphere from the layers in which they 
originally formed. As demonstrated in \cite{LimDust} this limiting case is 
appropriate for T~dwarfs which are discussed in this paper.
We stress that large uncertainties persist in the water opacities for parts
of the temperature range of this work \citep{2000ApJ...539..366A}. 

In addition to the opacity sources listed above and in \citet[and references
therein]{LimDust} the new LiCl line list presented in this paper has been added 
to our opacity database. In order to assess the effects of the new LiCl line 
data, we compare spectra calculated with and without this opacity source.
The models used in
the following discussion were all iterated to convergence for the parameters
indicated. The high resolution spectra which have the individual opacity
sources selected are calculated on top of the models. The LiCl line opacity 
data turned out to be too weak to influence the temperature structure 
of the atmosphere. 
The models have solar abundances with the non-depleted lithium abundance of 
log(n$_{\rm Li}$)=3.31.

\begin{figure}
\plotone{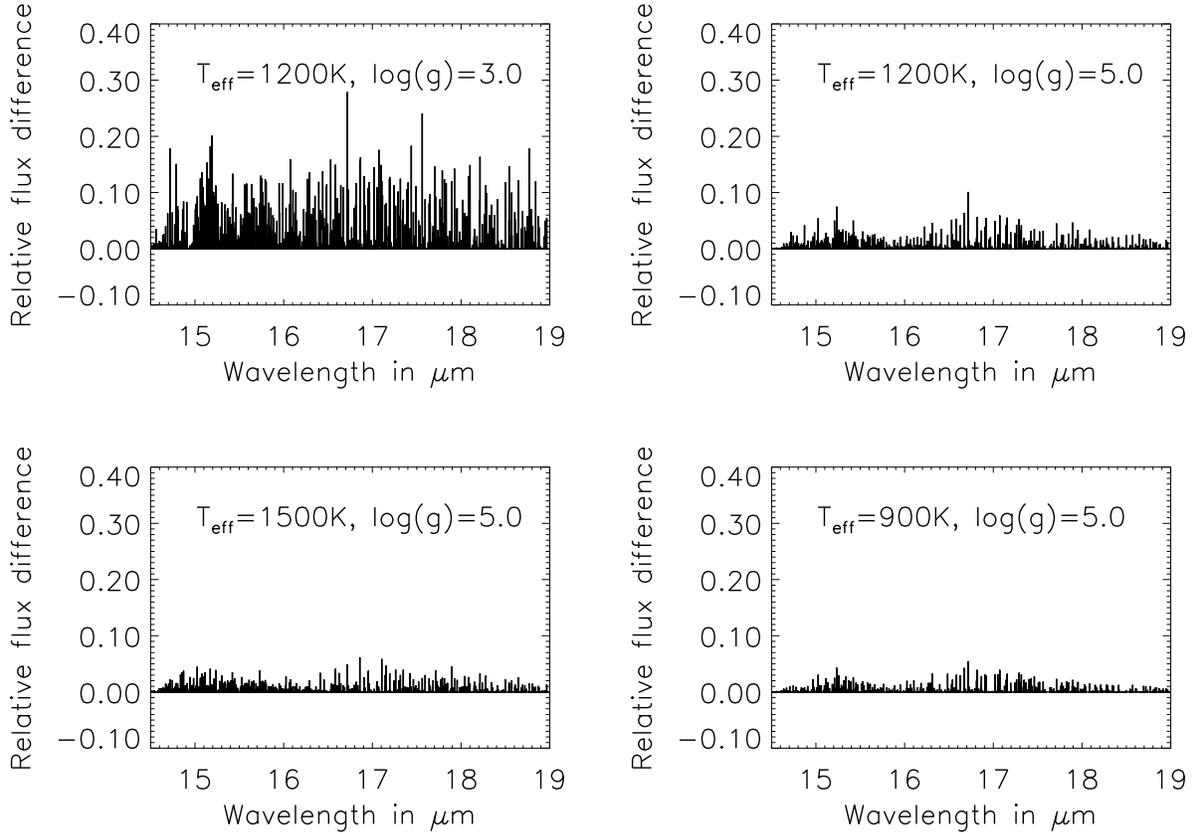}
\caption{Relative flux difference for brown dwarf model spectra with and 
without the present LiCl line opacity data for 
$T_{\rm eff}=900-1500\,$K, $\log(g)=3.0-5.0$ and solar abundances. 
The relative flux difference is defined here as $(F_0-F_{LiCl})/F_0$
where $F_0$ is the flux calculated with no LiCl absorption in our models and $F_{LiCl}$ 
is the flux including our new LiCl line list. The positive differences mean that there 
is less flux in the LiCl bands when including LiCl absorption data. 
The largest relative flux difference values correspond to the maximum effect of the 
LiCl absorption.  
\label{fig2}}
\end{figure}

We calculated models with log(g)=3.0, 4.0 and 5.0 and effective temperatures of
900~K, 1200~K and 1500~K which are typical parameters of old $(\log(g)=5.0, >1\mbox{Gyr})$ 
to young $(\log(g)=3.0, \simeq100~\mbox{Myrs})$ T~dwarfs. 
This parameter region turned out to be the one showing the strongest LiCl
features. As can be seen in Figure \ref{fig2} the effect of LiCl
is strongest for $T_{\rm eff}$=1200~K and log(g)=3.0 in the IR around the 
fundamental vibrational band origin at 15.8 $\mu$m and the relative flux difference 
is typically less than 20\% overall. The general strength of the 
LiCl absorption warrants inclusion in model calculations, but the lack of a distinct 
feature will make it hard to detect in an observed spectrum which is dominated by water
absorption.
However, the model parameters $T_{\rm eff}$=1200~K and log(g)=3.0 are particularly
interesting since these are parameters typical for very young (and hence very bright)
mid to early T~dwarfs.

\section{Conclusion}

Using an accurate hybrid potential and fully quantum-mechanical techniques, 
we have constructed a comprehensive and complete theoretical line list of spectroscopic 
accuracy for the $X~^1\Sigma^+$ electronic ground state of $^7$Li$^{35}$Cl.
Although LiCl appears to be a dominant Li-bearing gas over an extended domain 
of the $(T,P)$ diagram in cool dwarf atmospheres, synthetic spectra calculations with 
the stellar atmosphere code {\tt PHOENIX} suggest that flux differences resulting 
from the incorporation of this new line list are less than 20\% for parameters 
typical of young to old T~dwarfs. 
The strongest signature of LiCl for $T_{\rm eff}$=1200~K and log(g)=3.0 appears 
in the vicinity of the fundamental vibrational band origin at 15.8 $\mu$m, where 
the spectrum is dominated by water absorption. The current results
suggest that it will be difficult to measure the full inventory of elemental lithium 
in T~dwarfs after it is reposited into molecular species.

\acknowledgments

This work was supported by NASA grants NAG5-8425, NAG5-9222, and 
NAG5-10551 as well as NASA/JPL grant 961582, and in part by NSF grants 
AST-9720704 and AST-0086246 (and a grant to the Institute for Theoretical 
Atomic, Molecular \& Optical Physics, Harvard-Smithsonian CfA). Some of 
the calculations were performed on the IBM SP ``Blue Horizon'' of the San 
Diego Supercomputer Center, with support from the NSF, and on the IBM SP of 
NERSC with support from the DoE. This work also was supported in part 
by the P\^ole Scientifique de Mod\'elisation Num\'erique at ENS-Lyon. 
P.F.W. acknowledges ITAMP at Harvard University and SAO for travel support.



\clearpage

\clearpage

\onecolumn
\twocolumn


\clearpage

%
\begin{deluxetable}{crcccr} 
\tablecolumns{6} 
\tablewidth{0pc} 
\tablecaption{Line Oscillator Strengths and 
Transition Energies for Transitions in the $X~^1\Sigma^+$ 
state \label{tbl1}} 
\tablehead{ 
\colhead{} & \colhead{} &\colhead{} & \multicolumn{3}{c}{Transition Energy $(\mbox{cm}^{-1})$} \\ 
\cline{4-6} \\ 
\colhead{Band} & \colhead{Line} & \colhead{$f_{v'J',v''J''}$\tablenotemark{a}}  
& \colhead{Theory\tablenotemark{a}} & \colhead{Expt.\tablenotemark{b}} 
& \colhead{$\Delta E$\tablenotemark{c}}}    
\startdata 
$1\leftarrow 0$ & R(7) & 1.26217(-5) & 644.64 &  644.7380 & 0.0980\\
& R(10) & 1.21968(-5) & 648.35 &  648.4665 & 0.1165\\
& R(16) & 1.16041(-5) & 655.30 &  655.4703 & 0.1703\\
& R(19) & 1.13577(-5) & 658.55 &  658.7408 & 0.1908\\
& R(22) & 1.11270(-5) & 661.64 &  661.8543 & 0.2143\\ 
& R(26) & 1.08349(-5) & 665.53 &  665.7600 & 0.2300\\
& R(34) & 1.02818(-5) & 672.44 &  672.7089 & 0.2689\\
& R(49) & 9.30016(-6) & 682.23 &  682.5450 & 0.3150\\
& R(50) & 9.23624(-6) & 682.73 &  683.0491 & 0.3191\\
& R(58) & 8.72963(-6) & 686.06 &  686.3802 & 0.3208\\
\\
$2\leftarrow 1$ & R(1) & 3.25145(-5) & 628.00 &  628.0418 & 0.0418\\
& R(6) & 2.55717(-5) & 634.51 &  634.5921 & 0.0821\\
& R(12) & 2.38987(-5) & 641.78 &  641.9116 & 0.1316\\
& R(18) & 2.28181(-5) & 648.45 &  648.6304 & 0.1804\\
& R(23) & 2.20447(-5) & 653.55 &  653.7615 & 0.2115\\
& R(25) & 2.17516(-5) & 655.47 &  655.6910 & 0.2210\\
& R(28) & 2.13235(-5) & 658.21 &  658.4520 & 0.2420\\
& R(32) & 2.07686(-5) & 661.62 &  661.8852 & 0.2652\\
& R(36) & 2.02270(-5) & 664.75 &  665.0280 & 0.2780\\
& R(37) & 2.00932(-5) & 665.49 &  665.7710 & 0.2810\\
& R(48) & 1.86527(-5) & 672.38 &  672.6855 & 0.3055\\
\\
$3\leftarrow 2$ & R(5) & 3.89879(-5) & 624.51 &  624.5800 & 0.0700\\
& R(13) & 3.54663(-5) & 634.09 &  634.2288 & 0.1388\\
& R(16) & 3.46439(-5) & 637.42 &  637.5746 & 0.1546\\
& R(20) & 3.36623(-5) & 641.61 &  641.8000 & 0.1900\\
& R(23) & 3.29764(-5) & 644.58 &  644.7908 & 0.2108\\
& R(27) & 3.21029(-5) & 648.30 &  648.5320 & 0.2320\\
& R(40) & 2.94364(-5) & 658.45 &  658.7343 & 0.2843\\
& R(45) & 2.84505(-5) & 661.54 &  661.8405 & 0.3005\\
& R(46) & 2.82550(-5) & 662.10 &  662.4058 & 0.3058\\
& R(51) & 2.72851(-5) & 664.64 &  664.9493 & 0.3093\\
& R(52) & 2.70924(-5) & 665.09 &  665.3980 & 0.3080\\
\\
$4\leftarrow 3$ & R(12) & 4.75816(-5) & 624.26 & 624.3715 & 0.1115\\
& R(15) & 4.64316(-5) & 627.60 & 627.7262 & 0.1262\\
& R(25) & 4.32566(-5) & 637.63 & 637.8264 & 0.1964\\
& R(33) & 4.09981(-5) & 644.42 & 644.6672 & 0.2472\\
& R(38) & 3.96417(-5) & 648.10 & 648.3652 & 0.2652\\
& R(50) & 3.64842(-5) & 655.09 & 655.3840 & 0.2940\\
& R(58) & 3.44306(-5) & 658.27 & 658.5725 & 0.3025\\
\\
$5\leftarrow 4$ & R(17) & 5.70297(-5) & 621.13 & 621.2291 & 0.0991\\
& R(20) & 6.70073(-5) & 624.18 & 624.3003 & 0.1203\\
& R(32) & 5.14267(-5) & 634.85 & 635.0536 & 0.2036\\
& R(47) & 4.63923(-5) & 644.66 & 644.9326 & 0.2726\\
& R(55) & 4.37976(-5) & 648.23 & 648.5196 & 0.2896\\
\enddata 

\tablenotetext{a}{This work.} 
\tablenotetext{b}{\cite{jon87}.}
\tablenotetext{c}{$\Delta E=E_{Expt.}-E_{Theory}$}
\end{deluxetable} 


\begin{thebibliography}{}

\bibitem[\protect\citeauthoryear{Allard, Hauschildt, Alexander, Tamanai \&
  Schweitzer}{Allard et~al.}{2001}]{LimDust}
   Allard, F., Hauschildt, P.~H., Alexander, D.~R., Tamanai, A., 
   \& Schweitzer, A.  2001, 
   \apj, 556, 357
\bibitem[\protect\citeauthoryear{{Allard}, {Hauschildt} \& 
  {Schweitzer}}{{Allard} et~al.}{2000}]{2000ApJ...539..366A}
   {Allard}, F., {Hauschildt}, P.~H., \&  {Schweitzer}, A.  2000, 
   \apj, 539, 366
\bibitem[Barrado y Navascu\'{e}s, Stauffer \& Patten(1999)]{bar99} 
   Barrado y Navascu\'{e}s, D., Stauffer, J. R., \& Patten, B. M.  1999, 
   \apj, 522, L53
\bibitem[Basri, Marcy \& Graham(1996)]{bas96} 
   Basri, G., Marcy, G. W., \& Graham, J. R.  1996, 
   \apj, 458, 600
\bibitem[Brewer \& Brackett(1961)]{bre61}
   Brewer, L., \& Brackett, E.  1961, 
   Chem. Rev., 61, 425 
\bibitem[Bulewicz, Phillips \& Sugden(1961)]{bul61}
   Bulewicz, E. M., Phillips, L. F., \& Sugden, T. M.  1961, 
   Trans. Faraday Soc., 57, 921 
\bibitem[Burrows, Marley \& Sharp(2000)]{bur00} 
   Burrows, A., Marley, M. S., \& Sharp, C. M.  2000, 
   \apj, 531, 438
\bibitem[Burrows et al.(2002)]{bur02} 
   Burrows, A., Ram, R. S., Bernath, P., Sharp, C. M., 
   \& Milsom, J. A.  2002, 
   \apj, 577, 986
\bibitem[Chabrier, \& Baraffe(1997)]{cha97} 
   Chabrier, G., \& Baraffe, I.  1997, 
   \aap, 327, 1039
\bibitem[Cooley(1961)]{coo61} 
   Cooley, J. W.  1961, 
   Math. Computation, 15, 363
\bibitem[Dulick et al.(2003)]{dul03} 
   Dulick, M., Bauschlicher, C. W., Jr., Burrows, A., Sharp, C. M., 
   Ram, R. S., \& Bernath, P.  2003, 
   \apj, 594, 651
\bibitem[Hauschildt \& Baron(1999)]{hau99}
   Hauschildt, P. H., \& Baron, E.  1999, 
   J. Comput. App. Math. 102, 41
\bibitem[Huber \& Herzberg(1979)]{hub79} 
   Huber, K. P., \& Herzberg, G.  1979, 
   Molecular Spectra and Molecular Structure, Vol. IV, Constants 
   of Diatomic Molecules (New York: Van Nostrand Reinhold)
\bibitem[Jones \& Lindenmayer(1987)]{jon87} 
   Jones, H., \& Lindenmayer, J.  1987, 
   Chem. Phys. Lett., 135, 189
\bibitem[Lodders(1999)]{lod99} 
   Lodders, K.  1999, 
   \apj, 519, 793
\bibitem[Ogilvie(1992)]{ogi92} 
   Ogilvie, J. F.  1992, 
   Spectrosc. Lett., 25, 1341    
\bibitem[\protect\citeauthoryear{Partridge \& Schwenke}{Partridge \&
  Schwenke}{1997}]{ames-water-new}
   Partridge, H., \& Schwenke, D.~W.  1997, 
   \jcp, 106, 4618
\bibitem[\protect\citeauthoryear{Phillips \& Davis}{Phillips \&
  Davis}{1993}]{FeHberk2}
   Phillips, J.~G., \& Davis S.~P.  1993, 
   \apj, 409, 860
\bibitem[Rebolo, Martin, \& Magazzu(1992)Rebolo et al.]{reb92} 
   Rebolo, R., Martin, E. L., \& Magazzu, A.  1992, 
   \apj, 389, L83
\bibitem[\protect\citeauthoryear{Schwenke}{Schwenke}{1998}]{ames-tio}
   Schwenke, D.~W.  1998, 
   Chemistry and Physics of Molecules and Grains in Space,
   Faraday Discussion, 109, 321
\bibitem[Skory et al.(2003)]{sko03} 
   Skory, S. S., Weck, P. F., Stancil, P. C., \& Kirby, K.  2003, 
   \apj, 148, 599 
\bibitem[Thompson et al.(1987)]{tho87} 
   Thompson, G. A., Maki,  A. G., Olson, W. B., \& Weber, A.  1987, 
   J. Mol. Spectrosc., 124, 130 
\bibitem[Weck et al.(2003a)]{wec03a} 
   Weck, P. F., Schweitzer, A., Stancil, P. C., Hauschildt, P. H., 
   \& Kirby, K.  2003a, 
   \apj, 582, 1059 
\bibitem[Weck et al.(2003b)]{wec03b} 
   Weck, P. F., Schweitzer, A., Stancil, P. C., Hauschildt, P. H., 
   \& Kirby, K.  2003b, 
   \apj, 584, 459 
\bibitem[Weck et al.(2003c)Weck, Stancil \& Kirby]{wec03c} 
   Weck, P. F., Stancil, P. C., \& Kirby, K.  2003c, 
   \jcp, 118, 9997 
\bibitem[Weck et al.(2004)Weck, Kirby \& Stancil]{wec04} 
   Weck, P. F., Kirby, K., \& Stancil, P. C.  2004, 
   \jcp, 120, 4216

\end{thebibliography}
\end{document}